\documentclass[12pt]{pgarticle}
\usepackage{times}
\usepackage{graphicx}

\def\pBoxtimes{{\ooalign{\hfil\raise.2ex\relax
\hbox{$\times$}\hfil\crcr\hbox{$\Box$}}}}

\def\pBoxcirc{{\ooalign{\hfil\raise.2ex\relax
\hbox{$\circ$}\hfil\crcr\hbox{$\Box$}}}}

\def\pDiatimes{{\ooalign{\hfil\raise.2ex\relax
\hbox{$\times$}\hfil\crcr\hbox{$\Diamond$}}}}

\def\pBoxplus{{\ooalign{\hfil\raise.2ex\relax
\hbox{$+$}\hfil\crcr\hbox{$\Box$}}}}
\newenvironment{refs}{\bigskip 
   \begin{list}{}{\setlength{\leftmargin}{1cm}\setlength{\itemindent}{-1cm}
   \setlength{\topsep}{0cm}\setlength{\itemsep}{-0.12cm}} 
   \vspace*{-0.6cm}\footnotesize}{\end{list} }

\newcommand{\gtsim}{$\, \raisebox{-1.1mm}{\footnotesize $\stackrel{>}{\sim}$}
                    \,$}
\newcommand{\ltsim}{$\, \raisebox{-1.1mm}{\scriptsize $\stackrel{<}{\sim}$}
                    \,$}

\def\la{\mathrel{\mathchoice {\vcenter{\offinterlineskip\halign{\hfil
$\reset@font\displaystyle##$\hfil\cr<\cr\sim\cr}}}
{\vcenter{\offinterlineskip\halign{\hfil$\reset@font\textstyle##$\hfil\cr
<\cr\sim\cr}}}
{\vcenter{\offinterlineskip\halign{\hfil$\reset@font\scriptstyle##$\hfil\cr
<\cr\sim\cr}}}
{\vcenter{\offinterlineskip\halign{\hfil$\reset@font\scriptscriptstyle##$\hfil\cr
<\cr\sim\cr}}}}}

\def\ga{\mathrel{\mathchoice {\vcenter{\offinterlineskip\halign{\hfil
$\reset@font\displaystyle##$\hfil\cr>\cr\sim\cr}}}
{\vcenter{\offinterlineskip\halign{\hfil$\reset@font\textstyle##$\hfil\cr
>\cr\sim\cr}}}
{\vcenter{\offinterlineskip\halign{\hfil$\reset@font\scriptstyle##$\hfil\cr
>\cr\sim\cr}}}
{\vcenter{\offinterlineskip\halign{\hfil$\reset@font\scriptscriptstyle##$\hfil\cr
>\cr\sim\cr}}}}}

\def\cor{\mathrel{\mathchoice {\hbox{$\widehat=$}}{\hbox{$\widehat=$}}
{\hbox{$\reset@font\scriptstyle\hat=$}}
{\hbox{$\reset@font\scriptscriptstyle\hat=$}}}}

\def\fd{\hbox{$.\!\!^{\reset@font\r@mn{d}}$}}
\def\fh{\hbox{$.\!\!^{\reset@font\r@mn{h}}$}}
\def\fm{\hbox{$.\!\!^{\reset@font\r@mn{m}}$}}
\def\fs{\hbox{$.\!\!^{\reset@font\r@mn{s}}$}}

\def\farcm{\hbox{$.\mkern-4mu^\prime$}}
\def\farcs{\hbox{$.\!\!^{\prime\prime}$}}
\def\fp{\hbox{$.\!\!^{\reset@font\scriptscriptstyle\r@mn{p}}$}}

\newcommand{\Msun}{$M_{\odot}$}

\newcommand{\cM}{{\cal{M}}}

\begin{document}

\begin{center}
\Large\bf 
Space-Based UV/Optical Wide-Field Imaging and Spectroscopy: \\
Near-Field Cosmology and Galaxy Evolution Using Globular Clusters in Nearby Galaxies
\end{center}
\vskip 5pt

\begin{center}
{\bf Lead Author and Point of Contact:} \\
\medskip
Paul Goudfrooij \\  Scientist \\
Space Telescope Science Institute, 
3700 San Martin Drive, Baltimore, MD 21218 \\
Tel.: 410-338-4981; E-mail: goudfroo@stsci.edu \\ [1ex]

\begin{tabular}{l}
\multicolumn{1}{c}{\bf Co-Authors:} \\ [0.3ex]
Jean Brodie, University of California Santa Cruz, brodie@ucolick.org \\
Rupali Chandar, University of Toledo, Rupali.Chandar@utoledo.edu \\
Oleg Gnedin, University of Michigan, ognedin@umich.edu \\
Katherine Rhode, Indiana University, krhode@indiana.edu \\
Fran\c{c}ois Schweizer, Carnegie Observatories,
 schweizer@obs.carnegiescience.edu \\ 
Jay Strader, Harvard-Smithsonian Center For Astrophysics,
 jstrader@cfa.harvard.edu \\
Enrico Vesperini, Indiana University, evesperi@indiana.edu \\
Bradley Whitmore, Space Telescope Science Institute, whitmore@stsci.edu \\
Stephen Zepf, Michigan State University, zepf@pa.msu.edu \\
\end{tabular}
\end{center}

\begin{center}
\large\bf Abstract
\end{center}
\vspace*{-1mm}
\noindent
Star formation plays a central role in the evolution of galaxies and of the
Universe as a whole. Studies of star-forming regions in the local universe 
have shown that star formation typically occurs in a clustered fashion. Building a
coherent picture of how star clusters form and evolve is therefore critical to our 
overall understanding of the star formation process. Most clusters disrupt
after they form, thus contributing to the field star population. However, the
most massive and dense clusters remain bound and survive for a Hubble
time. These globular clusters provide unique observational probes of
the formation history of their host galaxies. In particular, the age and
metallicity can be determined for each globular cluster individually, allowing
the {\it distribution\/} of ages and metallicities within host galaxies to be
constrained.  

We show how space-based UV-to-Near-IR imaging covering a wide field of view
(\gtsim20$'$ per axis) and deep UV/Optical multi-object spectroscopy of 
globular cluster systems in nearby galaxies would allow one to place important
new constraints on the formation history of early-type galaxies and their
structural subcomponents (e.g., bulge, halo).    

\section{Globular Clusters as Fossil Records of the Formation History of
  Galaxies} 
\label{s:intro}

Infrared studies of star formation within molecular clouds have shown that
stars typically form in clusters or associations with initial masses
$\cM_{\rm cl,\,0}$ in the range $10^2 - 10^8$ \Msun\ (e.g., Lada \&
Lada 2003; Portegies Zwart et al.\ 2010). While most star clusters
with $\cM_{\rm cl,\,0}$ \ltsim $\,10^4$ \Msun\ are thought to disrupt
and disperse into the field population of galaxies within a few Gyr,
the surviving massive GCs constitute luminous compact 
sources that can be observed out to distances of several tens of
megaparsecs. Furthermore, star clusters represent the best known
approximations of a ``simple stellar population'', i.e., a
coeval population of stars with a single metallicity\footnote{Massive star 
clusters in the Milky Way host secondary populations with varying
relative light-element abundances (e.g., Gratton et al.\ 2012). However,
the effect of these variations to optical and near-IR colors is 
negligible (Sbordone et al.\ 2011).}, whereas the field stars
in galaxies typically constitute a mixture of populations.   
Thus, studies of globular cluster systems can constrain the {\it
  distribution\/} of stellar ages and metallicities whereas measurements of 
the integrated light of galaxies can only provide luminosity-weighted
averages of these key quantities. Consequently, globular clusters
represent invaluable probes of the star formation rate and chemical
enrichment occurring during the main star formation epochs within
their host galaxy's assembly history (see, e.g., reviews of Ashman \& Zepf
1998; Brodie \& Strader 2006).    

The study of extragalactic globular clusters was revolutionized by the
Hubble Space Telescope (HST). The main reason for this is that the size of
globular clusters is well-matched to diffraction-limited optical imaging with
a 2-m class telescope:\ a typical globular cluster half-light radius of
$\sim$\,3 pc at a distance of 15 Mpc corresponds to $\sim$\,0\farcs05 on the
sky, which is roughly the diffraction limit (and detector pixel size) in the
$V$ band for HST. This yields very high quality
photometry of globular clusters relative to ground-based optical 
imaging by beating down the high galaxy surface brightness in the central
regions of galaxies. Furthermore, it also allows robust measurements of
globular cluster radii, and hence of their dynamical status. 

Notwithstanding the important progress that HST imaging has facilitated in this
field, there is one critical property of globular cluster systems
that HST imaging {\it cannot\/} address well. Globular cluster systems around
massive early-type galaxies extend far into the galaxy halos, covering several
tens of arcminutes on the sky (e.g., Goudfrooij et al.\ 2001; Rhode \& Zepf 2001,
2004; Zepf 2005), while HST images only cover the central $\sim$\,3\farcm3 $\times$
3\farcm3. This is illustrated in Figure~1. Obviously, {\it wide fields of
  view\/} (\gtsim20$'$ per axis) are required to accurately
determine total properties of globular cluster systems (e.g., total numbers of
clusters per unit galaxy luminosity, color or metallicity distributions,
trends with galactocentric distance). Furthermore, the faint outer halos of
galaxies are thought to hold unique clues regarding the early assembly history
of galaxies, and bright globular clusters constitute one of the very few
probes that can be studied in these environments. In the following we
highlight a few key science questions in this growing field for which  new
space-based UV/Optical instrumentation can be expected to yield major steps
forward in our understanding of the formation and evolution of galaxies.

\vspace*{2.5mm}
\noindent
\begin{minipage}[t]{9.1cm}
{\includegraphics[width=9.cm]{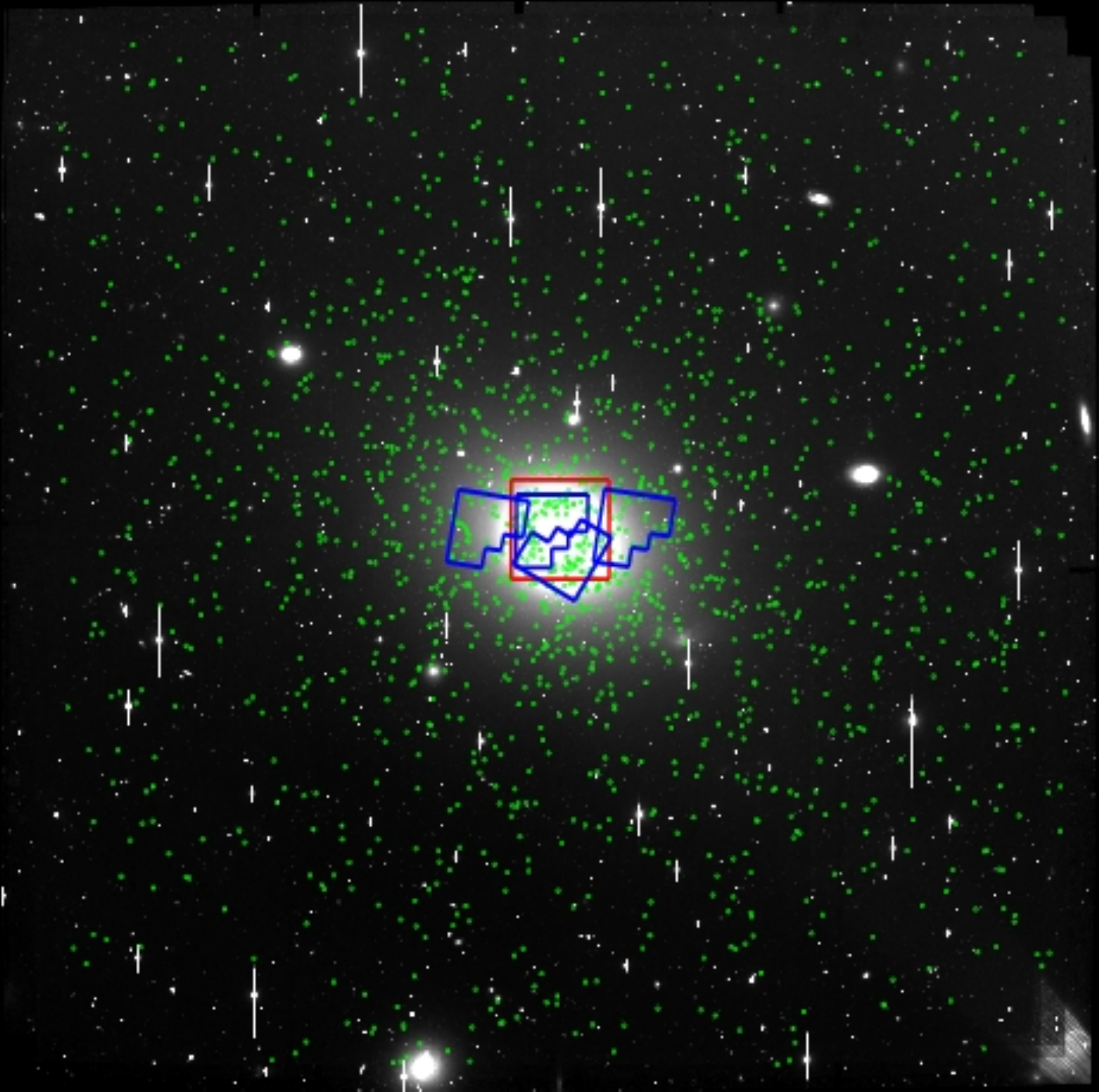}}
\end{minipage}
\hfill 
\begin{minipage}[b]{8.0cm}
{\baselineskip=0.95\normalbaselineskip 
{\bf Figure 1.} 
$R$-band KPNO 4-m/MOSAIC image of the giant elliptical galaxy NGC 4472 in the
Virgo cluster of galaxies, covering a 
$36' \times 36'$ field of view. Footprints of available HST/ACS and HST/WFPC2
images are drawn in red and blue, respectively. Globular cluster candidates
from Rhode \& Zepf (2001) are indicated as green dots. Note the small fraction
of globular cluster candidates covered by HST images, implying the need for
large and uncertain extrapolations when trying to extend conclusions from the
HST studies to the full systems of globular clusters. Figure taken from Zepf (2005). 
}
\vspace*{1.1cm}
\end{minipage}
\vspace*{-1ex}

\section{New Constraints on the History of Star Formation and Chemical Enrichment
  of Early-Type Galaxies}  
\label{s:SFR}

A key discovery of HST studies of globular cluster systems of luminous
galaxies was that their optical color distributions are typically bimodal
(e.g., Whitmore et al.\ 1995; Kundu \& Whitmore 2001; Larsen et al.\ 2001; Peng et al.\
2006). Figure 2 shows an example. 
Follow-up spectroscopy 
of bright globular clusters using 10-m-class telescopes has
indicated that both ``blue'' and ``red'' populations are typically old
(age\,\gtsim\,8 Gyr), implying that the color bimodality is mainly due to 
differences in metallicity (e.g., Cohen et al.\ 2003; Puzia et al.\ 2005). 
%
In broad terms, the metal-rich globular cluster population features colors,
metallicities, radial distributions, and kinematics that are similar to those 
of the spheroidal (``bulge'') component of early-type galaxies. In contrast,
the metal-poor globular cluster population has a much more radially extended
distribution, and is likely physically associated with metal-poor stellar
halos such as those found around nearby galaxies (e.g., Bassino et al.\ 2006;
Goudfrooij et al.\ 2007; Peng et al.\ 2008).     

\bigskip\noindent
\begin{minipage}[t]{8.8cm}
{\includegraphics[width=8.7cm]{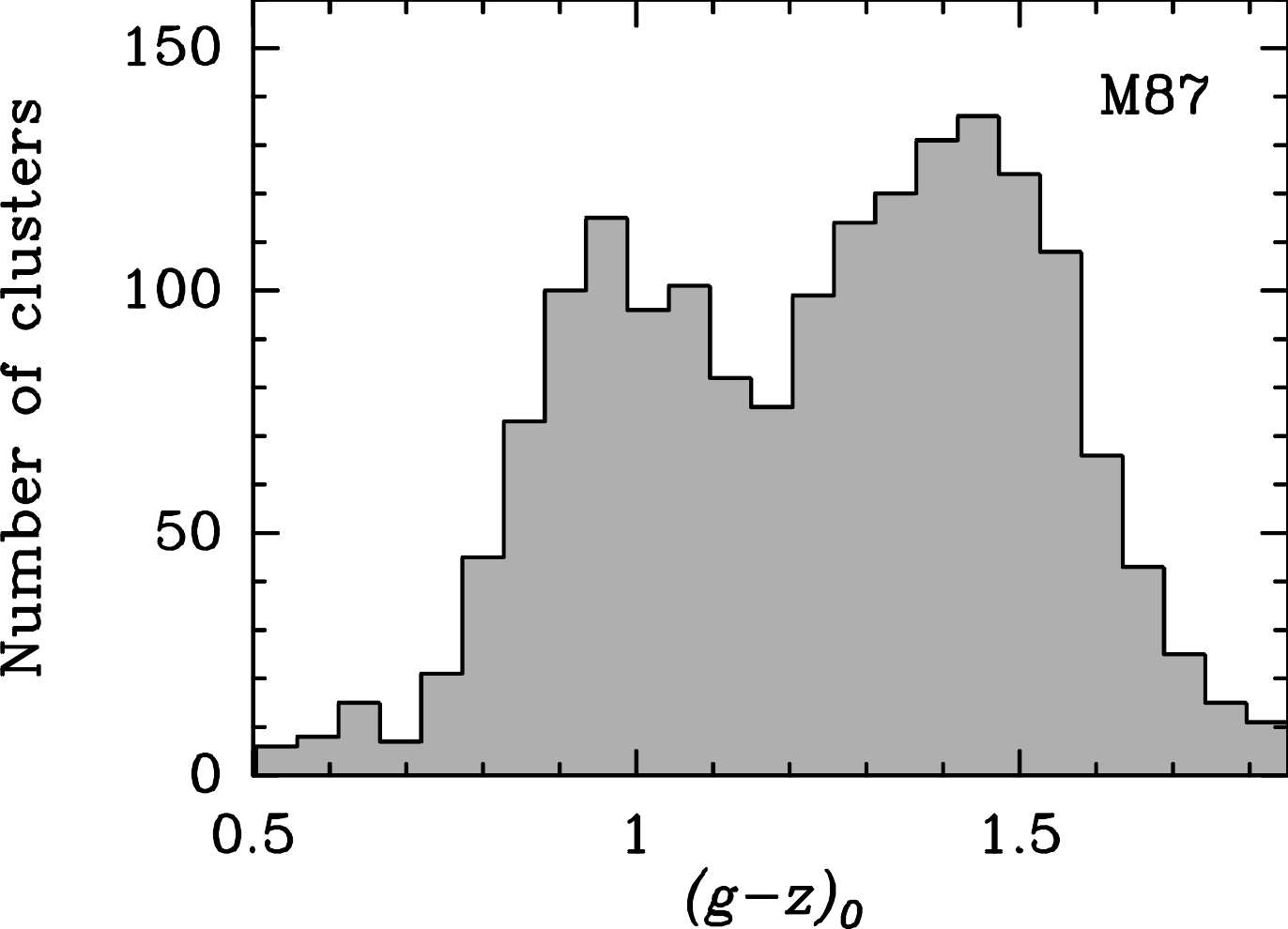}}
\end{minipage}
\hfill 
\begin{minipage}[b]{8.4cm}
{\baselineskip=0.95\normalbaselineskip 
{\bf Figure 2.} 
$g\!-\!z$ color distribution of globular clusters in the massive
elliptical galaxy M87 from Peng et al.\ (2006). Note the obvious color
bimodality, which has been confirmed to be mainly due to differences in
metallicity, and which is common among massive early-type galaxies in the
local universe.} 
\vspace*{1.3cm}
\end{minipage}
\vspace*{-0.ex}

\medskip
The bimodality in optical colors of globular clusters constitutes one of the
clearest signs that star formation in luminous early-type galaxies must have
been episodic.  However, we emphasize that the optical color
distributions do not significantly constrain {\it when\/} these events
occurred, or in what order. This is because optical colors alone cannot
generally distinguish between different combinations of age and metallicity
(the ``age-metallicity degeneracy''). A general understanding of the age and
metallicity distributions of globular cluster systems requires braking this
degeneracy. There are two primary and complementary ways to do this,
described below: 
\vspace*{-1.2mm}
\begin{enumerate}
\item {\bf The addition of near-infrared photometry to optical data}. The main power
  of this method (using color-color diagrams) is the ability to identify age
  differences (of order \gtsim\,25\% for high-quality data), due to the fact
  that near-IR colors are primarily sensitive to metallicity while optical
  colors are sensitive to both age and metallicity. 
  This approach resulted in the identification of substantial
  populations of intermediate-age metal-rich globular clusters in several
  early-type galaxies (Goudfrooij et al.\ 2001; Puzia et al.\ 2002; Hempel et
  al.\ 2007; Georgiev et al.\ 2012). The current limitation of this method is
  twofold. While HST has a powerful near-IR channel in its WFC3 instrument, its
  use is limited to the {\it innermost regions\/} of nearby galaxies
  due to its relatively small footprint of $\sim$\,2$' \times$ 2$'$ (cf.\
  Figure 1 above). The NIRCam instrument to be installed on the 6.5-m James Webb
  Space Telescope (JWST) will reach 2 mag fainter than HST in a given 
  integration time, but its footprint is similarly small. Conversely, while
  near-IR imaging instruments with reasonably large fields of view are
  starting to  become available on large ground-based telescopes
  (e.g., 7\farcm5 $\times$ 7\farcm5 for HAWK-I on the VLT),
  contamination of globular cluster candidate samples by compact
  background galaxies is a major concern for ground-based spatial
  resolution (see, e.g., Rhode \& Zepf 2001). As demonstrated by HST, 
  imaging at $\sim$\,0\farcs1 resolution effectively eliminates this concern
  due to the marginally resolved nature of globular clusters (cf.\ Section 1).  
  Thus, the study of galaxy formation and evolution by means of accurate
  globular cluster photometry will benefit tremendously from space-based
  wide-field UV/Optical imaging. A relatively simple multi-chip UV/optical
  camera installed on one of the two 2.4-m telescopes recently donated
  to NASA by the National Reconnaissance Office would be ideal for this (and
  many other) purpose(s). Their fast (f/1.2) primary mirror could easily
  yield a useful field of view of hundreds of square arcminutes per
  exposure at a resolution of $\sim$\,0\farcs1, providing accurate
  photometry of virtually {\it all\/} globular
  clusters associated with nearby galaxies with very little
  contamination. Along with a relatively standard suite of broad-band
  and narrow-band filters from the near-UV through the near-IR, such 
  an instrument would place important constraints on the formation and
  assembly history of massive early-type galaxies, particularly in
  their outer regions for which there currently are few other
  constraints. \\ [-3.5ex]  
\item {\bf Optical multi-object spectroscopy with large telescopes}. The main
  strength of this technique lies in the presence of 
  intrinsically strong absorption lines of several key elements in the optical
  region, which facilitates accurate determinations of overall
  metallicities and element abundance ratios that can be used to
  infer typical timescales of star formation  
  (e.g., Puzia et al.\ 2005, 2006). 
  However, this technique is currently only available from the ground and is
  therefore significantly hampered by the high surface brightness of the
  diffuse light of the inner regions of the host galaxies. In
  practice, this limits the application of this technique currently
  to mainly the {\it outer regions\/} of galaxies. This has caused a
  general lack of crucial spectroscopic information for the
  metal-rich globular clusters, which are located mainly in the inner
  regions. While future developments in the area of adaptive optics systems on
  large telescopes will enable high spatial resolution imaging and
  spectroscopy from the ground, they will do so only over a small (\ltsim
  1$'$) field of view which is not useful for spectroscopy of extragalactic
  star clusters. This science would however advance dramatically with a 8-m
  class  UV/Optical space-based telescope (such as the concepts proposed for
  ATLAST) equipped with a multi-object spectrograph with field of view of
  several arcmin per axis. Note that radial velocities resulting from such
  globular cluster spectra will also provide important kinematical probes in
  the outskirts of galaxies (where the diffuse light is too faint to 
  give useful information). 
  \\ [-4ex]    
\end{enumerate}

\section{From Star Clusters to the Field Star Population in Galaxies}
\label{s:disruption}

Star clusters begin disrupting (losing mass) as soon as they are
formed. Understanding how they do so as a function of cluster mass, time, and
environment is key to many questions in the study of star clusters and
their relation to galaxies. 
A main observable in this context is the star cluster mass function. HST
studies have shown that among young cluster systems in star-forming galaxies,
the mass function is well approximated by a power law ($\psi(M) \propto 
M^\alpha$  with $\alpha \simeq -2$, see, e.g., Fall et al.\ 2009 and
references therein). On the other hand, 
cluster mass functions in ancient galaxies such as giant early-type galaxies and
our Galaxy show a log-normal shape (e.g., Jord\'an et al.\
2007). This stark difference (illustrated in Figure 3) is most likely due to
dynamical evolution of the star cluster system. It is however not yet clear
how this important process happens in detail, and the recent literature
contains many different theoretical models and observational conclusions
regarding this transition. Being able to distinguish between the various ideas
will have relevant implications as to how, and to what extent, the field star
population in galaxies is built up over time from disrupting star clusters. 

\bigskip\noindent
\begin{minipage}[t]{8.4cm}
{\includegraphics[width=8.3cm]{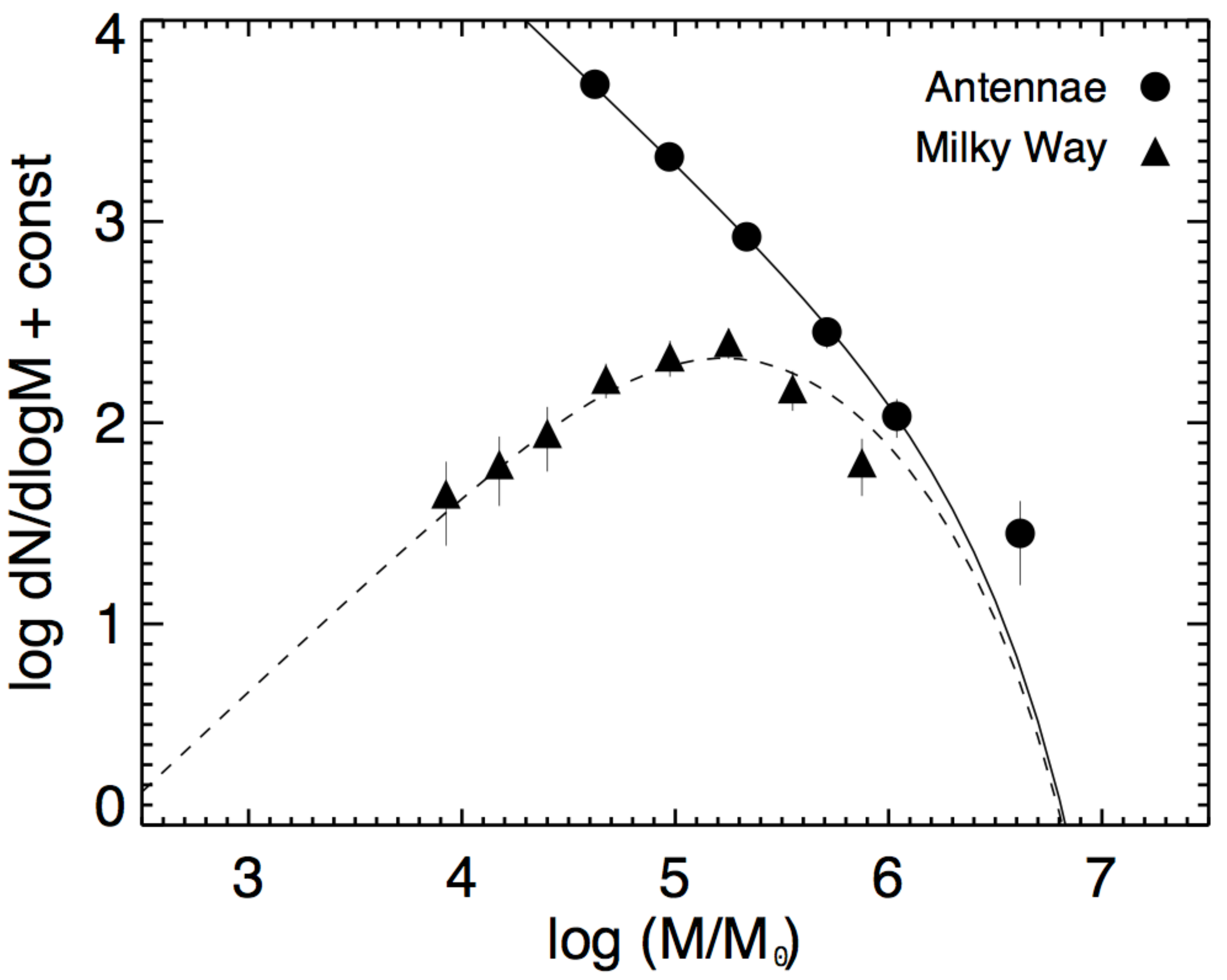}}
\end{minipage}
\hfill 
\begin{minipage}[b]{8.9cm}
{\baselineskip=0.95\normalbaselineskip 
{\bf Figure 3.} 
A comparison of the mass function of the young star cluster system of the
Antennae galaxies (Fall et al.\ 2009) with that of the globular cluster
mass function in the Milky Way. The stark difference between these mass
functions illustrates the important effect of dynamical evolution of star
clusters over time.}
\vspace*{1.1cm}
\end{minipage}
\vspace*{-0.1ex}

\medskip
As to the early stages of the cluster disruption process, the number of clusters per
unit log(age) in several star-forming galaxies appears to decline starting at
very young ages, suggesting that many clusters dissolve easily (e.g.,
Whitmore et al.\ 2007; Fall \& Chandar 2012). It is however not yet clear
which mechanism is most responsible for this rapid dissolution. Longer-term mass
loss of star clusters over a Hubble time is likely responsible for the
very different shapes observed for young and ancient star cluster systems
(cf.\ above). However, the disruption processes must also account for the
observation that the mass function of old globular clusters appears to be
similar among virtually all galaxies. Several different scenarios have been
proposed to explain this observation, each advocating different disruption 
mechanisms that act on different time scales (e.g., Vesperini \& Zepf 2003;
Parmentier et al.\ 2008; McLaughlin \& Fall 2008; Gieles et al.\ 2011). 

It is likely that the variety of proposed explanations for the difference
between mass functions of young and old star cluster systems is caused in
large part by the small footprint of HST images on the sky. In the central few
kpc of massive galaxies covered by HST images, the strong tidal field imposes
a relatively small range of mass densities on globular clusters in order for
them to survive tidal shocks for several Gyr (see, e.g., Gnedin 1997;
Goudfrooij 2012). This means that the current 
distribution of globular cluster sizes and mass densities (derived from HST
data) has no memory of the physical conditions occurring when the clusters
were formed, or even when they may have been accreted from dwarf galaxies (if
they were). This situation is quite different in the outer regions of
galaxies, where the tidal limit imposed by the galaxy potential on star
cluster sizes is much larger and observed star cluster sizes {\it do\/}
constrain the conditions occurring when the star clusters were formed or
accreted (e.g., Madrid et al.\ 2012). However, we simply do
not have adequate size information for star clusters in the outer regions of
massive galaxies at this time, and HST is not a suitable facility in this context.  

The determination of accurate ages for globular clusters at different
distances from the galaxy centers using space-based wide-field optical and
near-IR photometry and optical multi-object spectroscopy (cf.\ Section
\ref{s:SFR} above) will also yield important information to sort out the
relevance of various cluster disruption mechanisms (e.g., Goudfrooij 2012). 

In summary, accurate, deep cluster mass functions and size information for the
{\it full spatial extent of star cluster systems\/} will be key to our
understanding of dynamical evolution of star clusters and the nature of the
field star component in massive galaxies. Similar to the aforementioned study of
the star formation history of galaxies using globular cluster photometry (cf.\
Section \ref{s:SFR}), this study requires wide-field optical imaging 
with spatial resolution of order 0\farcs1 for which one of the 2.4-m space
telescopes donated to NASA by the National Reconnaissance Office would be a
very well-suited platform. 

\section{Concluding Remarks}

The increasing realization that the study of star clusters has direct
relevance for the basic processes involved in how galaxies assemble 
and evolve over time has placed this field at the forefront of
extragalactic research in recent years. We have described two fundamental 
questions that are of central importance in star cluster research, and
identified two types of future UV/Optical space telescope facilities that
would enable significant breakthroughs in these areas, providing important
new constraints for galaxy formation, assembly, and evolution. These two types
of facilities are:
\vspace*{-0.5mm}
\begin{enumerate}
\item A wide-field (of order 20$'$ $\times$ 20$'$), multi-detector imaging
  camera on a moderate-size space telescope with small focal ratio. One of the
  two f/1.2, 2.4-m space telescopes recently donated to NASA by the National
  Reconnaissance Office would be very well-suited to host such an
  instrument. \\ [-3.5ex] 
\item A 8-m class space telescope that includes a multi-object spectrograph
  that supports observations of \gtsim100 targets per exposure, covering a
  field of view of several arcmin per axis. The concepts proposed for ATLAST
  seem compatible with these requirements. \\ [-4ex]
\end{enumerate}

\bigskip
\noindent
{\bf References}

\noindent
\begin{minipage}[t]{3.4in}
\begin{refs}
\item Ashman, K. M., \& Zepf, S. E.\ 1998, {\it Globular Cluster
    Systems}, Cambridge University Press
\item Bassino, L. P., et al.\ 2006, A\&A, 451, 789
\item Brodie, J. P., \& Strader, J.\ 2006, ARA\&A, 44, 193
\item Cohen, J. G., Blakeslee, J. P., \& C\^ot\'e, P.\ 2003, ApJ, 592, 866
\item Fall, S. M., Chandar, R., \& Whitmore, B. C.\ 2009, ApJ, 704, 453
\item Fall, S. M., \& Chandar, R.\ 2012, ApJ, 752, 96
\item Georgiev, I. Y., Goudfrooij, P., \& Puzia, T. H.\ 2012, MNRAS, 420, 1317
\item Gieles, M., Heggie, D. C., \& Zhao, H.\ 2011, MNRAS, 413, 2509
\item Goudfrooij, P., et al.\ 2001, MNRAS, 328, 327
\item Goudfrooij, P., Schweizer, F., Gilmore, D., \& Whitmore, B. C.\ 2007,
  AJ, 133, 2737
\item Goudfrooij, P.\ 2012, ApJ, 750, 140
\item Gnedin, O. Y.\ 1997, ApJ, 487, 663
\item Gratton, R. G., et al.\ 2012, A\&AR, 20, 50
\item Hempel, M., et al.\ 2007, ApJ, 661, 768
\item Jord\'an, A., et al.\ 2007, ApJS, 171, 101
\item Kundu, A., \& Whitmore, B. C.\ 2001, AJ, 121, 2950
\end{refs}
\end{minipage}
\hfill
\begin{minipage}[t]{3.3in}
\begin{refs}
\item Lada, C. J., \& Lada, E. A.\ 2003, ARA\&A, 41, 57
\item Larsen, S. S., et al.\ 2001, AJ, 121, 2974
\item Madrid, J. P., Hurley, J. R., \& Sippel, A. C.\ 2012, ApJ, in press (arXiv:1208.0340)
\item McLaughlin, D. E., \& Fall, S. M.\ 2008, ApJ, 679, 1272
\item Parmentier, G., et al.\ 2008, ApJ, 678, 347
\item Peng, E. W., et al.\ 2006, ApJ, 639, 95
\item Peng, E. W., et al.\ 2008, ApJ, 681, 197
\item Portegies Zwart, S. F., et al.\ 2010, ARA\&A, 48, 431
\item Puzia, T. H., et al.\ 2002, A\&A, 391, 453
\item Puzia, T. H., et al.\ 2005, A\&A, 439, 997
\item Puzia, T. H., Kissler-Patig, M., \& Goudfrooij, P.\ 2006, ApJ, 648, 383
\item Rhode, K. L., \& Zepf, S. E.\ 2001, AJ, 121, 210
\item Rhode, K. L., \& Zepf, S. E.\ 2004, AJ, 127, 302
\item Sbordone, L., et al.\ 2011, A\&A, 534, A9
\item Vesperini, E., \& Zepf, S. E.\ 2003, ApJ, 587, L97
\item Whitmore, B. C., et al.\ 1995, ApJ, 454, L73
\item Whitmore, B. C., Chandar, R., \& Fall, S. M.\ 2007, ApJ, 133, 1067
\item Zepf, S. E.\ 2005, New A Rev., 49, 413
\end{refs}
\end{minipage}

\end{document}